\documentclass[sigconf]{acmart}
\AtBeginDocument{%
  \providecommand\BibTeX{{%
    \normalfont B\kern-0.5em{\scshape i\kern-0.25em b}\kern-0.8em\TeX}}}

\copyrightyear{2024}
\acmYear{2024}
\setcopyright{rightsretained}
\acmConference[CHI '24]{Proceedings of the CHI Conference on Human Factors in Computing Systems}{May 11--16, 2024}{Honolulu, HI, USA}
\acmBooktitle{Proceedings of the CHI Conference on Human Factors in Computing Systems (CHI '24), May 11--16, 2024, Honolulu, HI, USA}
\acmDOI{10.1145/3613904.3642059}
\acmISBN{979-8-4007-0330-0/24/05}

\usepackage{float} 
\usepackage{caption}
\usepackage{subcaption}
\usepackage{url}
\usepackage{cleveref}
\usepackage[labelfont=bf]{caption}
\usepackage{array}
\usepackage{multirow}
\usepackage{enumitem}    
\usepackage{graphicx}
\usepackage{caption}
\usepackage{tabularx}
\usepackage{csquotes}
\usepackage{soul}

\begin{document}

\title{Dissecting users' needs for search result explanations}



\author{Prerna Juneja}
\affiliation{%
  \institution{Seattle University}
  \country{WA, USA}}
  \email{pjuneja@seattleu.edu}

  \author{Wenjuan Zhang}
\affiliation{%
  \institution{Dataminr}
  \country{NY, USA}}
  \email{wzhang@dataminr.com}

    \author{Alison Marie Smith-Renner}
\affiliation{%
  \institution{Dataminr}
  \country{NY, USA}}
  \email{arenner@dataminr.com}

      \author{Hemank Lamba}
\affiliation{%
  \institution{Dataminr}
  \country{NY, USA}}
  \email{hlamba@dataminr.com}

      \author{Joel Tetreault}
\affiliation{%
  \institution{Dataminr}
  \country{NY, USA}}
  \email{jtetreault@dataminr.com}

      \author{Alex Jaimes}
\affiliation{%
  \institution{Dataminr}
  \country{NY, USA}}
  \email{ajaimes@dataminr.com}






\renewcommand{\shortauthors}{Juneja, et al.}

\begin{abstract} 
There is a growing demand for transparency in search engines to understand how search results are curated and to enhance users' trust. Prior research has introduced search result explanations with a focus on \textit{how} to explain, assuming explanations are beneficial. Our study takes a step back to examine \textit{if} search explanations are needed and \textit{when} they are likely to provide benefits. Additionally, we summarize key characteristics of helpful explanations and share users' perspectives on explanation features provided by Google and Bing. Interviews with non-technical individuals reveal that users do not always seek or understand search explanations and mostly desire them for complex and critical tasks. They find Google's search explanations too obvious but appreciate the ability to contest search results. Based on our findings, we offer design recommendations for search engines and explanations to help users better evaluate search results and enhance their search experience.
\end{abstract}



\maketitle

\section{Introduction}

{Search engines have become an integral part of our daily lives. According to a 2023 report, an overwhelming 93\% of all web traffic flows through search engines \mbox{\cite{SearchEn48:online}}. They not only facilitate information dissemination but also exert a profound influence on our beliefs, shape ideas, and mold the behaviors and perceptions of users \mbox{\cite{carroll2014search}}. For instance, a past study revealed that partisan bias in search engine rankings can substantially influence the voting preferences of individuals yet to make a decision \mbox{\cite{epstein2017suppressing,epstein2015search}}. Despite the power they hold, search engines have long operated as black boxes, providing little to no explanation about why specific search results are displayed to users and the order in which they appear.  This lack of transparency has led users to develop folk theories or incorrect concepts about search engine functioning. For instance, a previous study revealed that searchers mistakenly believe web search engines accept payment for higher-ranking positions in search results, beyond advertisements \mbox{\cite{thomas2019investigating}}. The lack of transparency has also been linked to limited engagement with search results \mbox{\cite{muramatsu2001transparent}} and has been shown to contribute to users' difficulty in finding desired information \mbox{\cite{thomas2019investigating,holman2011millennial,eslami2016first}}. Therefore, researchers and scholars have advocated for making search algorithms more accountable by being more transparent and interpretable. Studies have shown that by explaining how search works and communicating more information about search results, we can
help users be more efficient and effective searchers \mbox{\cite{ramos2020search,koenemann1996case,muramatsu2001transparent,hamilton2014path}} as well as raise user awareness
of biases in search engine results  \mbox{\cite{munson2013encouraging}}. }

{Despite the acknowledgment of the potential benefits of search explanations, 
there is limited work in this domain, often focusing on narrow aspects of search (such as query transformations), exploring how searchers perceive search engine decision-making, or testing the utility of specific types of search explanations. A notable gap exists in our understanding of what users think about search explanations---do users truly desire them, and if so, what specific needs and potential benefits are they seeking? This study aims to bridge the gap in understanding user perspectives on search explanations. By investigating when ``non-technical'' users (defined in Section \mbox{\ref{nontechparticipants}}) question search results, we identify situations where additional explanations or context could enhance their understanding. We explore what information search engines can provide to support informed decision-making and examine if and why users need search explanations for their various search needs. To navigate this inquiry effectively, we leverage the framework of search objectives identified by Rose et al. \mbox{\cite{rose2004understanding}}. This framework categorizes web searches into distinct goal-oriented categories, providing a structured approach to understanding and analyzing users' information-seeking behaviors. The framework includes categories such as \textit{directed-close searches}, where users seek a single, unambiguous answer to a specific question, \textit{advice searches}, aimed at acquiring guidance, ideas, suggestions, or instructions, etc. Our objective is to discern whether and why users find search explanations valuable in these specific scenarios, examining how search explanations contribute to the fulfillment of diverse search goals.}

In this study, we also assess the utility of current search engine explanation features. Popular web search engines like Google and Bing have traditionally offered brief text snippets with each search result, highlighting relevant keywords to indicate content alignment with queries. In response to the growing demand for transparency and accountability, modern search engines have recently introduced more sophisticated features designed to provide search explanations and additional context for search results. However, 
there is a limited understanding of users' awareness and the perceived effectiveness of these additional features. Our research 
 endeavors to bridge this gap by investigating users' perceptions of their usefulness. Overall, this study is guided by three fundamental research questions:

\begin{itemize}[leftmargin=*]
\item[] \indent \textbf{RQ1:}   Under what circumstances do users question the curation of search results? What additional information could facilitate their evaluation of the results?

\item[] \indent \textbf{RQ2:} For what search objectives do users find search result explanations helpful or unhelpful? In situations where explanations are deemed useful, what are the perceived benefits? What are the characteristics that users desire from search result explanations?

\item[] \indent \textbf{RQ3:} How do users perceive the effectiveness of the existing search result explanations and search result context provided by Google and Bing search engines?

\end{itemize}

To answer these research questions, we conducted a two-phase study. The first phase was an online survey aimed at screening participants and introducing them to various types of search objectives. The second phase was semi-structured interviews with 12 ``non-technical'' users, aiming to understand users' needs from search explanations. The study revealed that users tend to question search result curation when they encounter irrelevant results, inappropriate content (like ads or pornography), unknown sources, or have doubts about the selection of top-ranked results. Understanding the concept of explanations presented a challenge for certain participants, yet others found them useful in context-specific situations, especially for complex or high-stakes topics like medical searches and for searches where the objective is to find products or places they need to physically engage with. They prefer concise, actionable explanations that can aid in refining search queries and provide a better understanding of search result perspectives. In terms of their perception of existing search explanations, participants found Google's explanation too broad and obvious but appreciated the ability to provide feedback on search results and valued Bing's webpage preview feature. Our study reveals that users desire explanations not only to elucidate how search results are curated but also to gain additional insights about search results, all while maintaining agency over their search journey. Overall, our study delineates situations where search explanations are sought, outlines pivotal explanation attributes, and offers design recommendations for search engines.

\section{Related Work}

\subsection{Users' Interaction with Search Engines: Behavior, Decision-Making, and Mental Models}
Most studies have focused on understanding users' search behavior and search strategies by analyzing web log data. These studies reveal what people search on the web, how users formulate search queries, and patterns of query reformulations \cite{jansen2000real,spink2001searching,bilal2002differences,spink2001searching,bilal2002differences}. Studies have also delved into understanding how users decide to click and examine search results. An exploratory study conducted to determine how users find relevant search results found that factors like information present in the search result, users' beliefs, and time constraints under which the user is operating influence users' decisions \cite{barry1994user}. Similarly, another study identified several criteria used by users to assess the relevance of search results, such as content, the accuracy of information provided, users' understanding of the result (cognitive match), or
tasks, user beliefs (belief match), emotional response to the source (affective match) \cite{saracevic2007relevance}. Our work adds to this growing body of research. Rather than understanding what makes users click on a search result, we determine the situations in which users tend to question the results provided by a search engine and
identify elements that can enhance users' search experience.

Scholars have also examined users' mental models of online search. Studies in this area have sought to understand how users conceptualize search engines and the mechanisms behind them. Hendry and Efthimiadis, for instance, solicited sketches from students on how search engines work \cite{hendry2008conceptual}. The resulting figures contained a diverse array of concepts along with several misconceptions about how search engines work \cite{hendry2008conceptual}. Maramutsu et al. conducted a user study to understand users’ knowledge of and
reactions to the query transformations employed by search engines like stop word removals,  boolean operators, etc.  \cite{muramatsu2001transparent}. Zhang et al. studied undergraduate students' comprehension of various aspects of search engines, including search engine components, the search process, search result ranking mechanisms, and attributes of search engines \cite{zhang2008undergraduate}.
Thomas et al. combined surveys with interviews to understand how users conceptualize search engines' ranking algorithms \cite{zhang2008undergraduate}. The authors aimed to identify the concepts about how a search engine operates that users are familiar with and the concepts that are alien suggesting that search engines can add the concepts to the explanations about search results about which they have little understanding \cite{thomas2019investigating}. In contrast, our approach takes a different trajectory. Instead of pinpointing specific concepts to add to explanations, we investigate the necessity of explanations themselves and the situations in which they prove beneficial.

\subsection{Study context: explanations}
{The concept of explanation is multifaceted, drawing insights from various academic disciplines; however, there is no universally agreed-upon definition or formalization  \mbox{\cite{islam2020towards}}. According to psychologist Lombrozo, explanations serve as the ``currency in which we exchange belief''. They encompass both a cognitive process, wherein causes for an event, including specific counterfactual cases, are identified, and a resulting product---an explanation—derived from the cognitive explanation process \mbox{\cite{lombrozo2006structure}}. Numerous cognitive science theories emphasize that explanations frequently involve causal relations \mbox{\cite{lewis1986causal,srinivasan2021explanation}} and are supported by the idea that explaining an event entails offering information about its causal history. This notion aligns with the philosophical perspective that scientific explanations \mbox{\cite{salmon1984scientific,woodward2005making}} predominantly rely on causal relationships, wherein scientific facts are elucidated by appealing to their causes. Explanations are also thought of as similar to proofs in logic where a set of fundamental laws is presented as axioms, and the deductive sequences within the proofs form the explanation. \mbox{\cite{hempel1948studies}}. However, non-experts often struggle to understand these methods \mbox{\cite{srinivasan2021explanation}}.}

{Some definitions of explanation are specific to AI/ML and also take different forms. Some view it as any interpretable version of the original model \mbox{\cite{lundberg2017unified}}. Others define it as the set of features in the interpretable domain influencing a decision for a specific example \mbox{\cite{montavon2018methods}}. Additionally, a broader perspective encompasses all explanatory details about the model, including aspects like training data, performance, uncertainty, etc. \mbox{\cite{liao2020questioning}}. In the field of human-computer interaction, an explanation serves as a bridge between humans and a decision maker. It acts as a dual-purpose interface, accurately representing the decision maker while also being understandable to humans \mbox{\cite{guidotti2018survey}}. Scholars also highlight the social significance of explanations, going beyond the perspectives of developers and researchers. They define a `good explanation' as one that is not only understood by the explainer but also holds meaning for the person seeking it \mbox{\cite{brandao2019mediation}}. }

{In broader literature, the concept of explanation is also related to concepts of interpretability, transparency, trust, and fairness \mbox{\cite{abdul2018trends}}. Some scholars treat explainability and interpretability as synonymous, referring to how well a human can comprehend decisions in a given context \mbox{\cite{doshi2017roadmap}}. While some believe explainability encompasses a broader scope than interpretability \mbox{\cite{lipton2018mythos,habiba2022can,gilpin2018explaining}}.  
Interpretability is defined as the ability to summarize reasons for system behavior, gain user trust, or offer insights into decision causes. On the other hand, explainable AI goes beyond by being capable of defending actions, providing relevant responses, and allowing for audits \mbox{\cite{gilpin2018explaining}}. Privacy and security laws across countries have also aimed to define explanations. The European Union General Data Protection Regulation (GDPR), for instance, defines explanation as  ``meaningful information about the logic involved'' to individuals affected by automated decision-making systems \mbox{\cite{regulation2016regulation}}.  Despite the lack of consensus, all definitions share a common theme of emphasizing transparency and interpretability in providing comprehensive insights into decision-making processes. Taking inspiration from this theme, in this work, we adopt a broader definition of explanation. We define it as ``any feature or aspect that enhances the interpretability and transparency of a system, making it more understandable and clear to users''.}

\subsection{Search result explanations}
There exists a substantial body of scholarly work
dedicated to the design and generation of explanations
 (see \cite{abdul2018trends} for a detailed review). Previous research has focused on determining users' needs from explanations \cite{liao2020questioning,lim2009assessing,lim2009and}, determining the impact of explanations on user perceptions of and resulting interactions with these systems
\cite{kunkel2019let,cai2019effects,cheng2019explaining,dodge2019explaining,kocielnik2019will,kizilcec2016much,lim2009and,pu2006trust,cheng2019explaining,cramer2008effects},
as well as developing metrics to evaluate the quality and effectiveness of explanations \cite{hoffman2018metrics,carvalho2019machine}. However, most  of this research primarily focuses on the domains of AI and ML, where the  primary goal of explanations is to demystify how the model’s output is generated \cite{anik2021data,letham2015interpretable,ribeiro2016should,millecamp2019explain,wang2023pursuit,sendak2020human} or
within collaborative filtering and recommender
systems, where explanations shed light on the rationale behind the generation of specific recommendations \cite{cheng2019mmalfm,herlocker2000explaining,vig2009tagsplanations,tintarev2010designing}. {In contrast, studying explanations for search engines is still an under-explored area
in the broader field of explanations research.
This gap in research is particularly noteworthy given the unparalleled impact of search engines, facilitating access to an extensive array of information spanning educational research to health inquiries, setting them apart significantly from ML or recommendation systems that often focus on specific content domains. The impact of search engines is also evident with the results of recent surveys conducted by Edelman and Reuters that showed users trusting search engines
more than any other source including traditional news outlets and news recommended on social media  \mbox{\cite{2019Edel24:online,ReutersI82:online}}. Given their impact and concerns regarding potential biases \mbox{\cite{metaxa2021image, robertson2018auditing, hussein2020measuring, hu2019auditing, 2702520}}, scholars have begun advocating for transparent and
explainable search interfaces. }

{It is also crucial to note that, while both search engines and recommendation systems operate as black boxes, their operational distinctions are profound.    Search engines, propelled by explicit user queries, provide immediate and diverse results that hold the power to directly influence explicit decisions. In contrast, recommendation systems rely on implicit user behavior, gradually shaping decisions over time. 
Therefore transparency initiatives in the recommendation or AI/ML domains may fall short of adequately addressing the intricate interplay of user intent, query context, and the varied content types that search engines navigate. Thus, there is ongoing research focused solely on} generating explanations for search results \cite{yu2022towards,polley2021towards,baier2020explainable,ai2019explainable,ai2021model} and showing the impact of explanations on efficiency and trust \cite{ramos2020search,white2009characterizing,eickhoff2014lessons}, as well as explaining the relevance of search results to users \cite{coyle2005explaining,hearst1995tilebars}.  
However, there is still a limited understanding of if and when users find search explanations beneficial. Our study explores the potential advantages of search
explanations, identify the search objectives in which they prove valuable and ascertain the specific characteristics desired in valuable explanations.

\section{Method}
To answer our research questions, we employed two methods: online surveys\footnote{The surveys were administered using Qualtrics software.} followed by semi-structured interviews. 
As part of the online surveys, we first administered a screening survey to determine eligibility for participation. Those meeting the criteria were invited to complete the study survey and participate in the interviews. {Our study was reviewed for ethical concerns by our institution.} We describe our recruitment strategy and method details of our study below.

\subsection{Recruitment}
\subsubsection{Platforms} We recruited participants for our study from two platforms: Prolific\footnote{https://www.prolific.co/} and Upwork\footnote{https://www.upwork.com/}. Prolific is a crowdsourcing website similar to Amazon Mechanical Turk, while Upwork is a platform that connects employers with freelancing professionals. To ensure compliance with Prolific's community guidelines, we created a separate study for screening purposes constituting both surveys. Users who qualified were added to the allowlist of the main interview study. On the other hand, Upwork allows the inclusion of a pre-screening survey within the job post itself. Therefore, we designed a single job post on Upwork with two milestones: a study survey and an interview. Upwork also offers the functionality to target professionals with specific criteria. For our study, we specifically targeted native or bilingual English speakers with professionals having job success scores of at least 90\%. We excluded participants with skill sets in Development \& IT. We paid participants  \$5 to fill out the study survey and \$20 to participate in the interview.

\subsubsection{Targeted user group}  \label{nontechparticipants} 
{In our study, our primary focus is on comprehending the perspectives of ``non-technical'' individuals. Drawing on the work of Thomas et. al. \mbox{\cite{thomas2019investigating}}, we define non-technical users as those lacking a computer science/IT background and not possessing a deep familiarity with concepts related to Artificial Intelligence (AI), Machine Learning (ML), and information retrieval (IR). Participants were recruited based on self-reported data obtained through a screening survey (as detailed in Section \mbox{\ref{screen}}). We hypothesized that such individuals would exhibit a strong understanding of information retrieval concepts, heightened awareness of search engine functionalities, and familiarity with the principles of explainability and interpretability as well as awareness of information disorders. It is crucial to note, however, that our conceptualization of the non-technical group is inherently limited, and we acknowledge this constraint in Section \mbox{\ref{limitation}}.}

\begin{table*}[]
\begin{tabular}{m{0.4cm}|m{4.2cm}|m{4cm}|m{5cm}}
\textbf{No.} & \textbf{Search objective} & \textbf{Example search queries} & \textbf{Perceived helpfulness of explanations} \\ \hline
1 & \textit{Directed-closed:} You wanted to get an answer to a question that has a single, unambiguous answer & US 2022 midterm election date, 401k contribution limit 2022 & \includegraphics[width=0.3\textwidth]{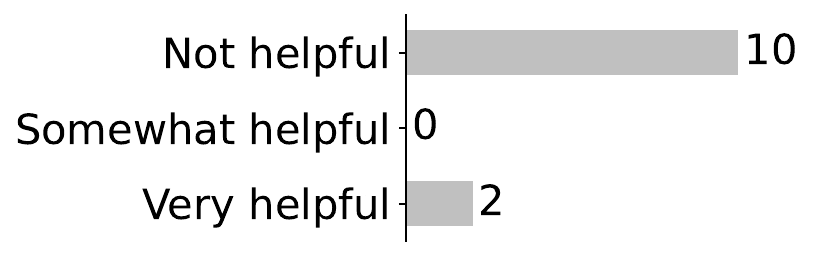}   \\ \hline
2 & \textit{Directed-open:} You wanted to get an answer to an open-ended question, or one with unconstrained depth & Fun things to do in Seattle, date ideas Philly &  \includegraphics[width=0.3\textwidth]{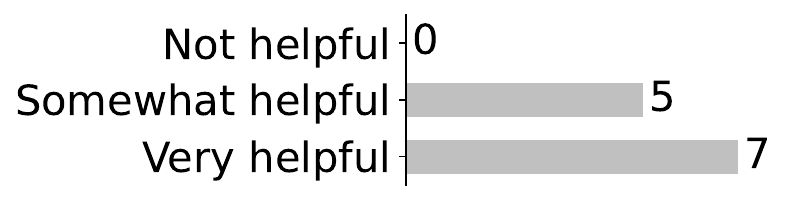} \\ \hline
3 & \textit{Undirected:} You wanted to learn anything/everything about a topic & Russia-Ukraine war, platelet disorders & \includegraphics[width=0.3\textwidth]{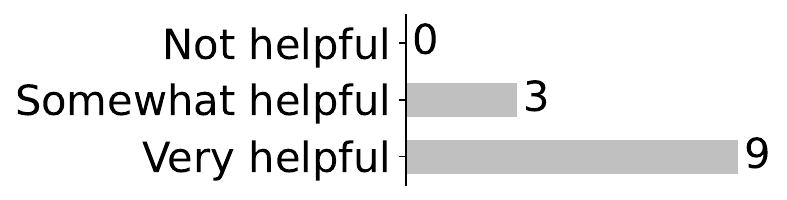}  \\ \hline
4 & \textit{Advice :} You wanted to get advice, ideas, suggestions, or instructions & How to bake a chocolate cake, how to change state drivers license &  \includegraphics[width=0.3\textwidth]{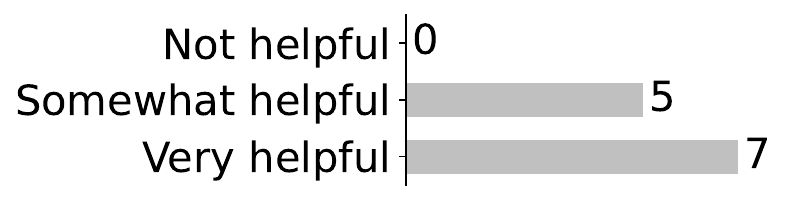} \\ \hline
5 & \textit{Locate:} Your goal was to find out whether/where some real world service or product can be obtained & Where to buy a kayak, handyman services near me & \includegraphics[width=0.3\textwidth]{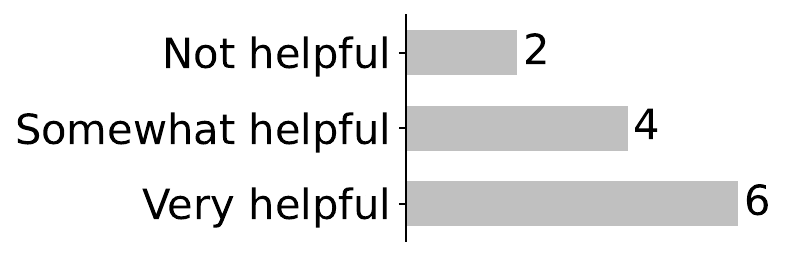}  \\ \hline
6 & \textit{List:} Your goal was to get a list of websites, each of which might be candidates for helping me achieve some underlying, unspecified goal & Top US universities, best ways to learn investment & \includegraphics[width=0.3\textwidth]{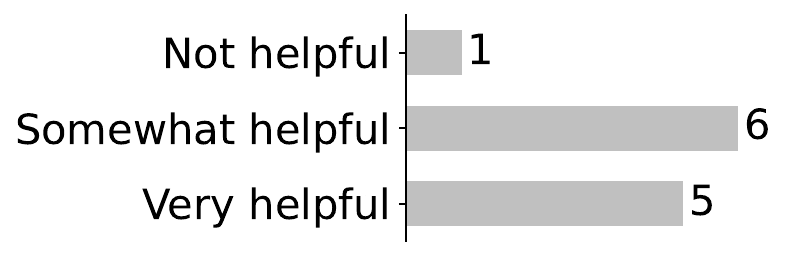} 
\end{tabular}
\caption{Table displays the search objectives along with two sample search queries: one provided by us as an example and the other filled in by participants during the survey. Additionally, the table presents the count of participants who found explanations helpful for each search objective.}
\label{search_scenarios}
\Description[Search objectives]{Table displays the search objectives along with two sample search queries: one provided by us as an example and the other filled in by participants during the survey. Additionally, the table presents the count of participants who found explanations helpful for each search objective.}
\end{table*}

\subsection{Online surveys}

\subsubsection{Screening survey} \label{screen}
Participants needed to meet the following criteria to be eligible: 1) be at least 18 years old, 2) 
reside in the United States, 3)
 use search engines at least once or twice a day, and 4) possess a novice to intermediate level of expertise in AI and concepts of IR\footnote{We disqualified participants who indicated intermediate or expert-level expertise in AI and IR. We also excluded individuals with degrees in computer science, information technology, and library or information science as these backgrounds could indicate a higher level of expertise in the aforementioned concepts.}. To ensure data quality, we added an attention check question and disqualified participants who answered it incorrectly. Additionally, we added a text box question asking participants to describe how search engines like Google, Bing, etc. provide search results from billions of possibilities\footnote{The exact phrasing of the question was borrowed from \cite{thomas2019investigating}}. We clarified that the question was not a test of participants' capabilities and we did not expect perfect answers. Participants who provided off-topic, spam, or responses were disqualified from further participation. 
 While our initial intention was to disqualify participants who provided sophisticated answers, including references to concepts such as Tf-Idf, Latent Semantic Analysis, Query Expansion, etc., it's noteworthy that none of the participants provided responses of such complexity. Out of the 86 users who completed the screening survey (51 from Prolific and 35 from Upwork), 29 participants qualified for the study (19 from Prolific and 10 from Upwork).

\subsubsection{Study survey}

Our study aims to provide valuable insights into users' needs for search explanations across a wide range of search objectives. By incorporating these objectives, we achieve a more comprehensive understanding of users' expectations and preferences when seeking information through search engines in various real-life situations. We adopted the hierarchy of search objectives outlined in \cite{rose2004understanding}, with a specific focus on informational searches where the goal is to seek information about the query topic rather than navigational searches where the goal is to reach the homepage of a particular website or institution. Table \ref{search_scenarios} provides an overview of the six search objectives identified by the authors. In the study survey, participants were introduced to these objectives, and provided with detailed explanations and examples. They were then prompted to recall and share examples of search queries from their own past browsing history that were relevant to each search objective. Additionally, we sought participants' perspectives on the potential helpfulness of search engine explanations for each search objective, encouraging them to consider the context of the examples they provided. To ensure the attentiveness of participants, we included two attention-check questions in the survey. Participants who answered either of these questions incorrectly were subsequently disqualified from further participation in the study.  Out of the 25 participants who qualified, 12 individuals successfully participated in the subsequent interview study (3 from Prolific and 9 from Upwork).

\begin{table*}[]
\begin{tabular}{lllll}
\hline
\textbf{ID} & \textbf{Gender} & \textbf{Age} & \textbf{Education} &  \textbf{Ethnicity}\\ \hline
P1 & M & 25-34 & Bachelor & Other (Armenian)\\
P2 & F & 25-34 & Master & White/Caucasian\\
P3 & F & 18-24 & Bachelor & White/Caucasian\\
P4 & M & 35-49 & High school & White/Caucasian \\
P5 & F & 18-24 & Bachelor & White/Caucasian\\
P6 & M & 18-24 & Bachelor & Hispanic or Latino \\
P7 & M & >=50 & Bachelor & White/Caucasian\\
P8 & M & 35-49 & Bachelor & White/Caucasian\\
P9 & M & 18-24 & High school & White/Caucasian\\
P10 & F & >=50 & High school & American Indian or Alaska Native\\
P11 & F & 35-49 & Bachelor & Mixed \\
P12 & F & 35-49 & High school & Black or African-American
\end{tabular}
\caption{Demographics information of our interview participants}
\label{participants}
\Description[Participant demographics]{Demographics information of our interview participants.}
\end{table*}

\begin{figure*}[t]
 \centering
    \begin{subfigure}[b]{0.54\textwidth}
        \centering    
        \setlength{\fboxsep}{1pt} 
    \setlength{\fboxrule}{0.5pt}
    \fbox{\includegraphics[width=1\textwidth]{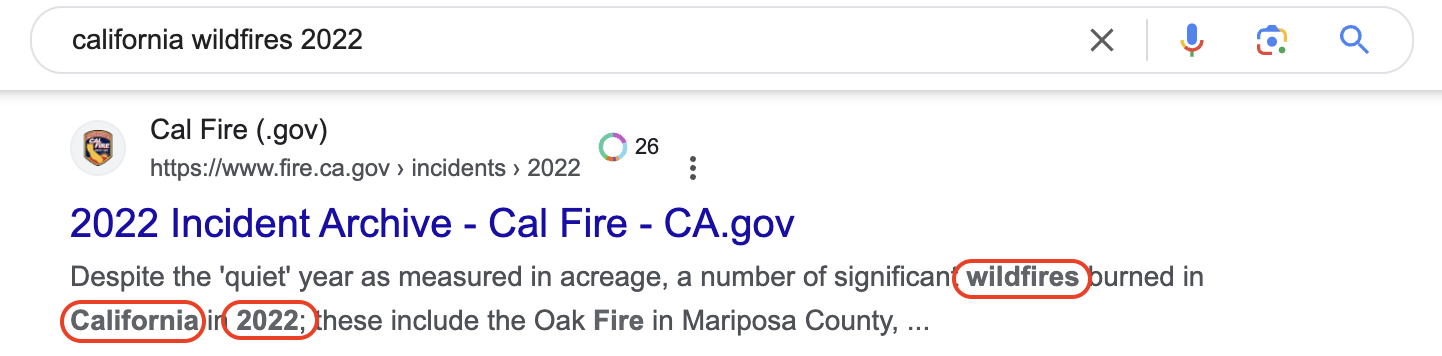}}
        \caption{}
        \label{snippet}
        \Description[Snippet of Google search result]{Figure showing highlighted keywords in the snippet of text accompanying the search result, offering insights into the alignment of the search result with the user's search query.}
    \end{subfigure}\\
    \begin{subfigure}[b]{0.8\textwidth}
        \centering    
        \setlength{\fboxsep}{1pt} 
    \setlength{\fboxrule}{0.5pt}
    \fbox{\includegraphics[width=1\textwidth]{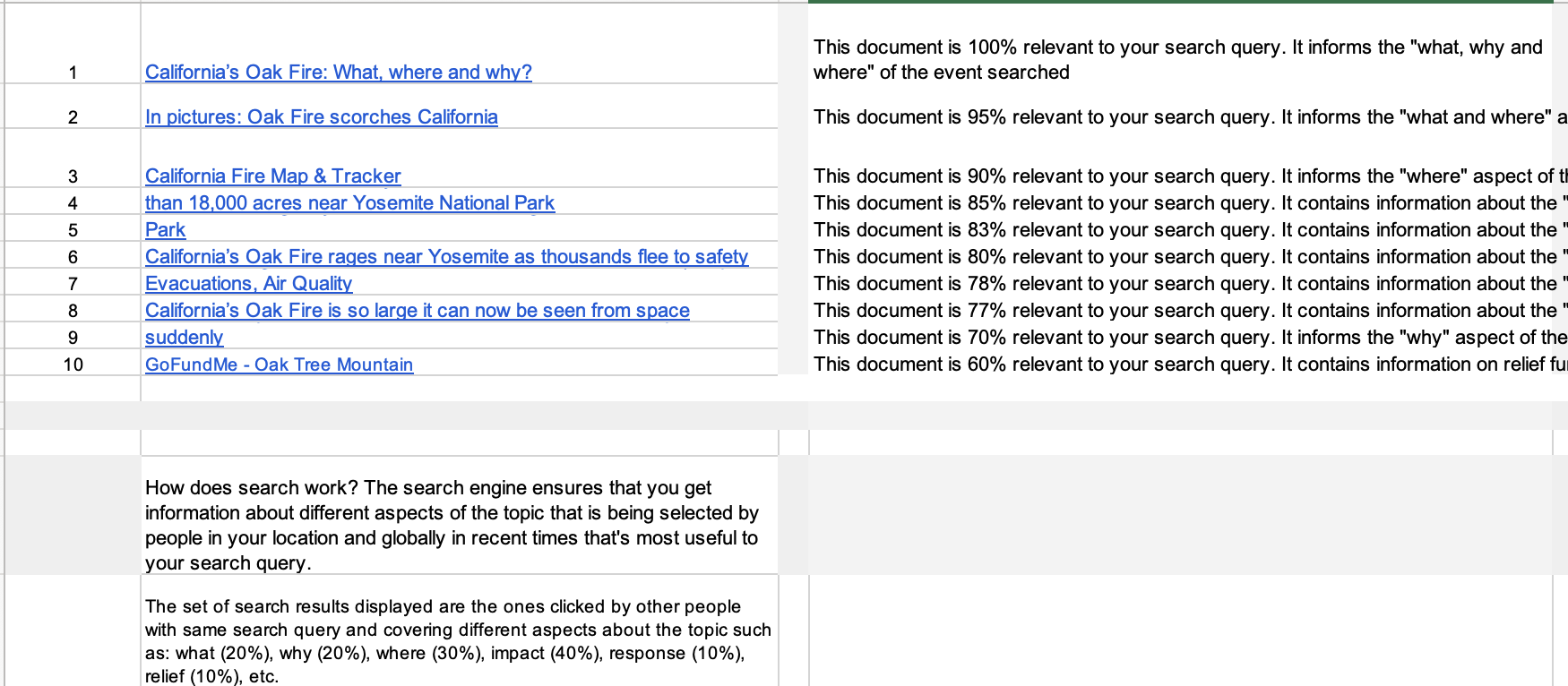}}
        \caption{}
        \label{example}
        \Description[Sample search explanation]{Figure showing search explanations provided as an example during the interview. This example illustrates an overall explanation describing how the search engine works and, in addition, explains how relevance influences the ranking of search results.}
    \end{subfigure}
    \caption{(a) Figure showing highlighted keywords in the snippet of text accompanying the search result, offering insights into the alignment of the search result with the user's search query, (b) Figure showing search explanations provided as an example during the interview. This example illustrates an overall explanation describing how the search engine works and, in addition, explains how relevance influences the ranking of search results. }
\end{figure*}

\subsection{Semi-structured interview}

\subsubsection{Interview participants}

Table \ref{participants} displays the basic demographic information of the participants.
The participant group consisted of an equal number of males and females, ensuring gender balance. Additionally, participants from various age groups were represented in the study. The majority of participants identified as White/Caucasian, and a significant proportion held a Bachelor's degree or equivalent qualification.

\subsubsection{Interview protocol}
Participants were first asked general warm-up questions, including the frequency of their search engine usage and whether they had encountered search explanations on the search results page. They were then prompted to recall situations where they questioned the search results returned by the search engines. Next, we conducted a contextual inquiry task, wherein participants were instructed to search for a specific query on their preferred search engine and share their screen. They were asked to describe their process for selecting search results and provide suggestions for additional information that could enhance their search experience. {For this task, we selected two search queries voluntarily provided by participants for various search scenarios in the study survey. These included queries about both non-controversial topics such as `best ways to learn investment,', `date ideas in Philly', etc. and more controversial subjects (if such examples were provided by the participants)  such as `covid-19 vaccines,' `climate change', etc. Recall that participants had initially selected these search queries from their browser histories during the study survey, mirroring real-world scenarios where they actively formulate search inquiries based on their information needs. Our goal was to ensure that the chosen queries resonated with participants' familiarity and interests. 
During the interview process, some participants chose to search for additional or different search queries in addition to the ones we selected for them. We allowed this flexibility to ensure participants' comfort and autonomy. Note that the objective of this task is not to determine specific search features needed in particular scenarios but rather to capture the range of potential features that could enhance participants' overall search experience.}

The interview then shifted to discussing the need for search explanations in various search objectives presented in the study survey. {Our initial step involved introducing participants to the concept of search explanations, which presented a twofold challenge. On one hand, our objective was to sidestep any inadvertent biases that could influence participants and lead them toward pre-existing notions about explanations. On the other hand, we recognized that non-expert users might find it challenging to reflect on something unfamiliar. We addressed this by first providing participants with our definition, explaining, ``When you submit a query to a search engine, the results you receive are carefully chosen from countless possibilities. Any feature or aspect that enhances the interpretability and transparency of the system, making it more understandable to you, can be considered a search explanation.''}

{We then proceeded to illustrate examples of what these features or aspects could encompass. Initially, we highlighted snippets (Figure \mbox{\ref{snippet}}), the succinct summaries provided by search engines, often bolding relevant keywords to give users an indication of the content's alignment with their query. Following this, we showcased the current explanation format offered by Google (Figures \mbox{\ref{g1}} and \mbox{\ref{g2}}). We expanded on the concept by explaining that search explanations can take various forms. They can offer an overall description of how the search engine operates, elucidate why a specific search result or set of results was or wasn't presented, clarify the reasoning behind the ranking order of search results, and provide insights into why alternative ranking orders might be absent. Furthermore, explanations can shed light on the absence of potential websites or documents from search results. To ensure participants grasped these concepts, we documented these types in a Google document and allowed participants time to review and absorb the information. For those seeking additional clarity, we shared a Google sheet with dummy search results for the query `California wildfires 2022' along with corresponding example search explanations (refer to Figure \mbox{\ref{example}} for an illustration). We selected this topic due to its relevance in the news during the time of the interview study, and its uncontroversial nature.}

After introducing the concept of explanations, we showed participants the search objectives and their study survey responses. They were then asked to elaborate on why explanations would be helpful or not helpful for each objective and specify the information they needed from the search engine to understand the returned search results. Finally, participants were presented with the search results pages of Google and Bing, including any existing explanations provided by the platforms. They were then requested to provide feedback on the effectiveness of the explanations and identify areas that could be improved. This activity aimed to gather valuable insights into participants' perspectives on the strengths and weaknesses of the current explanation features provided by these platforms, helping to inform potential enhancements.

{During the interviews, we implemented various measures to foster a non-judgmental and comfortable atmosphere for the participants. We provided the option for participants to turn off their video feed if they preferred (however, they were required to share their screens during the session). They were actively encouraged to engage in discussions on topics they felt comfortable with, focusing the conversation on search queries they had previously disclosed in the survey to ensure familiarity with the subject and enhance their comfort in sharing. Additionally, we aimed to facilitate open dialogue and understanding, refraining from probing or questioning personal beliefs.}

All interviews were conducted by the first author in September 2022 using Lookback\footnote{https://www.lookback.com/}, with participants providing their consent for audio and video recording. In addition to the recorded interviews, the first author took detailed notes during the discussions.  The interviews lasted for approximately 60-70 minutes.  

\subsubsection{Data analysis} 
The first author independently reviewed the observation notes and transcripts from the interviews, making necessary corrections to the text. Guided by the three research questions, a qualitative analysis was conducted using an inductive thematic analysis approach \cite{guest2011applied}. The first author performed two rounds of open coding to identify and develop themes, which are reported in the subsequent sections.  We refer to participants as P1-12, using gender-non-specific pronouns such as ``they'' and ``them''.


\section{RQ1 Results: Understanding Users' Search Behavior and Identifying Opportunities to Enhance their Search Experience}

\subsection{Situations prompting users to question the search results curated by search engines}
We prompted participants to think about instances where they questioned the search results curated by search engines. Our aim was to identify situations where users would seek additional context or explanations about the results returned by search engines.  Four key themes emerged from the interviews. 

\subsubsection{Irrelevant Search Results:} Several participants mentioned their tendency to question the logic behind search result curation when encountering irrelevant or undesirable information in response to their queries. In such situations, participants usually reword their search queries to get more relevant search results. P11 explains, \textit{`sometimes like if I do like a search ..and.. I'll see things that are really unrelated to what I'm searching for..then I'll question..and usually what I'll do is I'll reword what I'm searching for ''}.

\subsubsection{Unknown and Unrecognized Sources:}  Participants also raised concerns when search results were returned from unknown or unrecognized sources. As P8 noted, ``\textit{I usually [question search results] from web pages that are not like a name that I recognize}''. Ads also drew attention, with participants noting that they ``\textit{don't think too deeply about the search results except in case of some questionable sponsored [links] or the ads}'' (P2). Participants also revealed that they usually question sponsored content when it ``\textit{gets pushed to the top of the search results page}'' (P5). 

\subsubsection{Inappropriate Content: } One participant mentioned questioning the functioning of search engines when inappropriate content is returned. They explained, ``\textit{Sometimes when I search for something innocent, the search engine may present results containing explicit content, which makes me uncomfortable. In such cases, I would prefer to see more family-friendly content}'' (P12). 

\subsubsection{Questioning Top Results:} Participants also expressed their curiosity about the order in which search results were displayed, particularly why certain outcomes were given priority over others. For example, P1 elaborated, \textit{``let's say I would search up basketball players and it would bring up a whole bunch of results that have that basketball player keyword or whatever..But as far as the order and how it populated that..why this player...I question that ''}.

\subsection{Users' perspectives on how to enhance search experience }
During the interviews, participants provided valuable insights into how search engines could enhance the search experience and facilitate informed decision-making about which results to click on. Their suggestions revolved around several key aspects:

\subsubsection{Website Previews:} Participants expressed a strong desire for website previews that would allow them to gain insights into the content before clicking on a link. As one participant mentioned, a website preview could save time by showing relevant information upfront, stating, ``\textit{Maybe if I'm hovering over a link or a website, it would give me a preview of where I'm about to go... it might save me a little bit of time so I know which website I should visit}'' (P1). On similar lines, another participant highlighted they want previews like  YouTube snippets where ``
 \textit{there was an image and you can hover over it and like a little box can pop up and it could be just like the beginning of the article}'' (P5).

\subsubsection{Detailed Result Descriptions:} Participants also expressed the need for improved descriptions accompanying the search results. They suggested that more detailed descriptions would help them make informed decisions about clicking on a result. P7 stated that they would like access to an expanded description by having ``\textit{an option for an additional sentence or two to further assess relevance''} of the search results (P7). 

\subsubsection{Visual Elements:}
Participants emphasized the importance of visual elements, such as images, in assessing the content's relevance. They believed that visual cues would help them \textit{``quickly determine whether a result aligns with [their] search intent''} (P12).

\subsubsection{Credibility Signals:} Participants also emphasized the importance of credibility signals such as website views, ratings, and source information. One participant likened search engines to social media platforms, stating, ``\textit{It's hard for me not to think of it like as a social media type thing where you can see how many views there are or how many people could recommend the website as valuable or reliable... I don't know like a certificate or something}'' (P3). P8 further added---``\textit{What I'm looking for is this a trusted source... And I know because it's ranked on Google here, it probably is, but I don't know why... if it's because a lot of people go to the result}''
One participant expressed a preference for logos, specifically mentioning the desire to see recognizable logos like the CNN logo, as a quick visual cue to identify trusted websites (P6). Another participant indicated the importance of publishing date as a credibility indicator, expressing the need to know ``\textit{when the article was actually made and put on the internet, how recent was this written}'' (P3). \\

\noindent It is important to note that during the interviews, participants did not spontaneously express a need for search result explanations suggesting that, in their natural search behavior, users may not inherently seek or expect such explanations. Instead, their focus is primarily on features that indicate credibility and enhance search efficiency.

\section{RQ2 Results: User Perspectives on the Effectiveness of Search Result Explanations }

To address the second research question, we introduced the concept of search explanations to participants. Additionally, we presented them with their survey responses and prompted them to elaborate on their answers by explaining why they believed explanations would be either helpful or not helpful in each objective. Table \ref{search_scenarios} presents the search objectives.  We encouraged them to use the example search queries they provided for each search objective to contextualize their responses. It is important to note that some participants found the concept of explanations challenging to comprehend, even after being presented with examples. In fact, two participants struggled to grasp the concept of explanations entirely. In this section,  we explore the search objectives in which participants perceived search results as useful and not useful.

\subsection{Utility of explanations in various search objectives} 
As evident from Table \ref{search_scenarios}, for all search objectives except the first, the majority of participants found search explanations helpful. While participants describe explanations serving different purposes in different objectives (Section \ref{rq2c}), we found that the need for explanations is also highly context-dependent. Participants sought explanations for tasks that require ``\textit{deep engagement}'' or when ``\textit{they have the mental capacity and time}'' (P11) to delve into the search topic. In contrast, for day-to-day searches, participants never ``\textit{second-guessed the search results}'' (P4) and thus,  did not feel the need for explanations. Participants highlighted the importance of more detailed explanations in high-stakes topics such as medical-related searches, indicating a desire to know if reputable sources were being presented. However, for opinion-based topics, participants considered explanations to be less significant, as their focus was on acquiring information that aided in making personal decisions. As participant P6 stated, ``\emph{For how to vote by mail, you want like a government site or something...if you wanted advice on stuff like blood pressure you would want to uh you would definitely want to know why you are getting shown a website... for how to bake a cake, I wouldn't really care to know exactly how they decided to show me the best recipe}''.

Furthermore, participants highlighted that for searches where the goal is to \textcolor{darkgray}{\textbf{find real-world service or product}}, explanations become particularly valuable as participants anticipate interacting with these products or visiting specific places. As P3 expressed---``\textit{[Explanations] would be very helpful because I actually have to go to those places... I actually am gonna be physically somewhere or involved with this product... so explanations would be really helpful}''.

Participants generally found explanations unnecessary when seeking \textcolor{darkgray}{\textbf{a single, unambiguous answer}}.  For such searches, search engines often provide featured snippets or answer boxes, which are information boxes displayed at the top of search engine results pages (SERPs) to directly answer user queries without the need to click on any search results. The simplicity of such queries led participants to believe that explanations would only clutter the website and divert users' attention. 
 P6 explains, ``\textit{For search goal like that.. like midterm election date, Prime Minister, stuff like that... where it's just a clear answer, there's no debate as to what the answer is..if an explanation was added, I feel like it would sort of clutter the website in a way and just kind of distract me instead}''. Trust in the search engine's ability to provide accurate results for such search scenarios was also highlighted, with one participant mentioning, ``\textit{I'm not really super concerned about where that definition comes from, as long as it's from a trusted source and correct...which I would imagine all the top search results for Google on that search term would be correct.}'' (P4).

\begin{table*}[]
\renewcommand{\arraystretch}{1.5}
\begin{tabular}{m{2cm}m{3cm}m{6cm}}
 \hline
\multicolumn{3}{c}{\vspace{1mm}\textbf{Perceived benefits of search results explanations}} \\ \hline

\multicolumn{1}{m{3.5cm}|}{\textbf{Themes}} & \multicolumn{1}{l|}{\textbf{Search objectives}} &\textbf{Example quote} \\ \hline

\multicolumn{1}{m{3.5cm}|}{Increase trust in search results} & \multicolumn{1}{m{4.5cm}|} {\begin{itemize}[left=0em]
\item Get answer to an open-ended question
\item  Learning everything about a topic
\item Getting advice, ideas, or instructions
\item  Obtaining real-world service or product
\item  Get a list of websites
\end{itemize}} & ``\textit{There are so many different ways to get scammed online....it would be good to have some kind of  reassurance from Google or Bing like, hey this is a trusted source or  the more products when I'm looking for pricing or to buy an item is it's kind of the better for me}''---P8 \\ \hline

\multicolumn{1}{m{3.5cm}|}{Assist in selecting search results} & \multicolumn{1}{m{4.5cm}|}{\begin{itemize}[left=0em]
\item Get answer to an open-ended question
\item  Learning everything about a topic
\item Getting advice, ideas, or instructions
\item  Obtaining real-world service or product
\item  Get a list of websites
\end{itemize}} & ``\textit{Explanations could be useful in and uh in the decision-making process also be helpful maybe in picking which sites to open and explore further, if they had that kind of uh explanation}''---P7 \\ \hline

\multicolumn{1}{m{3.5cm}|}{Help in guiding and refining the search} & \multicolumn{1}{m{4.5cm}|}{\begin{itemize}[left=0em]
\item Get answer to an open-ended question
\item Learning everything about a topic
\item Getting advice, ideas, or instructions
\item  Obtaining real-world service or product
\item Get a list of websites
\end{itemize}
} & ``\textit{The explanations can help you understand your own thought process a little bit more..can help you understand better from seeing those options, understand where and choose where you want to go from there versus like high level at the beginning, When you're first typing in your original search}''---P2 \\ \hline

\multicolumn{1}{m{3.5cm}|}{Understand the  perspective and categories shown in search results} & \multicolumn{1}{m{4.5cm}|}{\begin{itemize}[left=0em]
\item Get answer to an open-ended question
\item  Learning everything about a topic
\end{itemize}} &  ``\textit{So having an explanation as to why certain search results pop up more than others might help me like save time or figure out like, okay, like this is maybe a new technique I didn't know or this is a new perspective I had never thought about when tackling this issue}''---P5 \\ 
\end{tabular}
\caption{Table providing a summary of the perceived benefits of search result explanations as expressed by the participants along with an example quote.}
\label{benefits}
\Description[Perceived benefits of search explanations]{Table providing a summary of the perceived benefits of search result explanations as expressed by the participants along with an example quote.}
\end{table*}

\subsection{Perceived benefits of search results explanations for various search objectives}\label{rq2c}
In this section, we present the perceived benefits of search explanations identified through the interviews for various search objectives. Table \ref{benefits} provides a summary.

\subsubsection{Explanations increase trust in search results:}
Participants believed that search explanations had the potential to bolster trust by offering insights into the credibility of search results across various search objectives.  As one participant explained, ``\textit{So having an explanation about like why was this so highly ranked... like okay, this is like a reputable source, so it's maybe if it's like a journal, it's peer-reviewed or experts or well-known people in the field are cited here, those kinds of things would be much helpful}'' (P5). Such credibility signals allow participants to \textit{``differentiate between [results].. and determine which are more trustworthy than others''} (P6). 
Participants also indicated explanations could increase trust by mentioning if the search result is \textit{``paid content or sponsorship and is there not necessarily because of relevance, but because of money''} (P5). 

 \begin{small}
\begin{displayquote}
\enquote{I would like to see why I'm getting certain searches and certain results back and if something's kind of boosted to get its way up there because if I'm learning about a topic I don't wanna have to read 10 different articles to make sure what I'm reading is correct, I wanna be able to click on something, read through it, have some sense of reliability and uh and kind of move on.} - P8
\end{displayquote}
\end{small}

\subsubsection{Explanations could serve as decision-making tools by assisting in the selection of search results:} Participants expressed that explanations could assist with the decision-making process about picking which sites to open and explore further.

 \begin{small}
\begin{displayquote}
\enquote{Explanations could be useful in the decision-making process and helpful in picking which sites to open and explore further, if they had those kinds of explanations, it could indicate that the items are pulled from major online marketers and listed in highest to lowest or lowest to highest cost, that kind of things could be useful.
 } - P7
\end{displayquote}
\end{small}

\begin{table*}[]
\renewcommand{\arraystretch}{1.5}
\begin{tabular}{m{5cm}m{9.5cm}}
 \hline
\multicolumn{2}{c}{\vspace{1mm}\textbf{Desired characteristics of search explanations}} \\ \hline
\multicolumn{1}{l|}{\textbf{Themes}} & \textbf{Example quote} \\ \hline
\multicolumn{1}{m{5cm}|}{Explanations should be an easy read} & ``\textit{When you're using a search engine, your key interest is hit and run type of a thing. You want to know what you want to know as quickly as possible. You want to spend time with the information and not try to find the information…So to be really helpful explanations would need to be a very quick read}''---P7. \\ \hline
\multicolumn{1}{m{5cm}|}{Explanations should be actionable} & ``\textit{Explanations could tell the searcher what they're looking for and sort of help sort the searches..If they can have a little button where you can change the ranking based on possibility..like what is closest, where can I find cars in the 15-20000 dollar range or 50 to \$60,000 range?}''---P7 \\ \hline
\multicolumn{1}{m{5cm}|}{Explanations should indicate reputability, trustworthiness, and popularity of the source} & ``\textit{So having an explanation about like Okay, why was this so highly ranked…Also if it's like okay, we this is like a reputable source, so it's maybe if it's like a journal, it's peer-reviewed or experts or well-known people in the field are cited here, Those kinds of things would make me would be much more helpful in a search like that}''---P5 \\ \hline
\multicolumn{1}{m{5cm}|}{Explanations should indicate viewpoints} & ``\textit{If I want to learn something, I would want to know who is teaching me this thing that I'm trying to learn and make sure that their views aligned with mine, so there will be no uh kind of disparity in our viewpoints}''---P1 \\
\end{tabular}
\caption{Table providing a summary of the characteristics desired by the participants in search explanations.}
\label{characteristics}
\Description[Desired Characteristics of search explanations]{Table providing a summary of the characteristics of search explanations that are desired by the participants.}
\end{table*}

\subsubsection{Explanations can help in guiding the search process and refining the queries:} Participants highlighted that explanations could assist in understanding their own thought processes and guide them toward more relevant search results. They mentioned that searches where the \textcolor{darkgray}{\textbf{goal is either to get an answer to an open-ended question}} or \textcolor{darkgray}{\textbf{learn everything about a topic}} could benefit from explanations that help users comprehend the different options and make informed decisions about their search direction. 

 \begin{small}
\begin{displayquote}
\enquote{Yeah, I think that would be helpful because if your question is open-ended or can be pretty deep and go a bunch of different ways…..The explanations can help you understand your own thought process a little bit more..can help you understand better from seeing those options, understand where and choose where you want to go from there versus like high level at the beginning When you're first typing in your original search, you probably don't know those broader or those more specific uh levels of that at that point in time, otherwise you would have searched for something a bit more specific, to begin with
 } - P2
\end{displayquote}
\end{small}

Participants also suggested the idea of using explanations to categorize or classify information to help users find specific details or narrow down their search for the aforementioned search objectives. Such categorization would also allow them to quickly identify redundant information and focus on acquiring new knowledge, thus saving time and making the search efficient. 

 \begin{small}
\begin{displayquote}
\enquote{You're asking the open-ended question coz you need as much information as possible...in such case, explanations should sort the searches by category..so you start with general and then you get down to specific category} - P7
\end{displayquote}
\end{small}


 \begin{small}
\begin{displayquote}
\enquote{I could hover [over the explanations]... I might be like, okay, but I know X, Y, and Z already, so I'm gonna skip that... but now I'm looking here and this has a new piece of information, and I jump in... having an explanation would save time and also help you narrow down what you don't know.
 } - P5
\end{displayquote}
\end{small}

Participants also believed that explanations could enable them to \textit{``refine [their] search terminology and search terms to obtain results that were more specific to their situation''} (P9). P3 further elaborated that search engines could achieve this by suggesting the use of Boolean operators, for example, asking users to  ``\textit{add one word, subtract one word, or change something}'' (P3) in the search query.

\subsubsection{Explanations help in understanding the perspectives or categories shown in the search results:} The interviews revealed that explanations could play a vital role in enhancing users' understanding of the perspectives and viewpoints presented in the search results, especially for searches where the \textcolor{darkgray}{\textbf{goal is either to get an answer to an open-ended question}} or \textcolor{darkgray}{\textbf{learn everything about a topic}}.  As P4 mentioned, ``\textit{having an explanation might make me think more about it like, 'Oh, so this perspective is not being shown or this website is not being shown because maybe the search engines found it not being appropriate or not being uh something and a viewpoint is being hidden or suppressed.}'' 

Moreover, explanations can lead users to discover new approaches and viewpoints when approaching a problem. P5 emphasizes this point, stating that ``\textit{having an explanation as to why certain search results pop up more than others might help me...figure out like this is maybe a new technique I didn't know or a new perspective I had never thought about.}'' By providing insights into the rationale behind the prominence of specific results, explanations offer users the opportunity to broaden their horizons and consider alternative perspectives or techniques. P1 further stated that they would want explanations to tell them ``\textit{which side is the information coming from?}'' to enable them to assess potential biases or agendas in the information presented to them by the search result.



\begin{figure*}[]
    \centering
    \vspace{1cm}
    \begin{subfigure}[b]{0.54\textwidth}
        \centering    
        \setlength{\fboxsep}{0.3pt} 
    \setlength{\fboxrule}{1pt}
    \fbox{\includegraphics[width=1\textwidth]{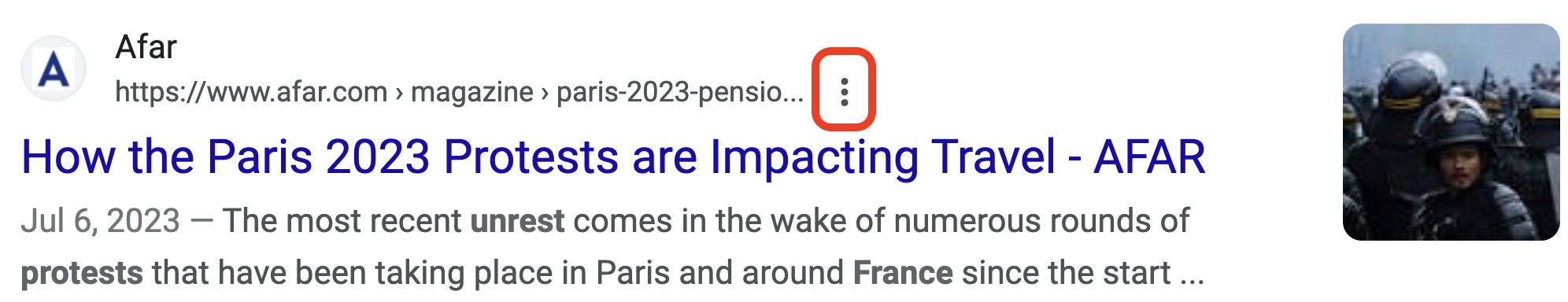}}
        \caption{}
        \label{g_dots}
        \Description[Google search]{The search explanation provided by Google can be accessed by clicking on the three dots present along each search result as shown in the figure.}
    \end{subfigure}
    \begin{subfigure}[b]{0.54\textwidth}
        \centering    
        \setlength{\fboxsep}{0.3pt} 
    \setlength{\fboxrule}{1pt}
    \fbox{\includegraphics[width=1\textwidth]{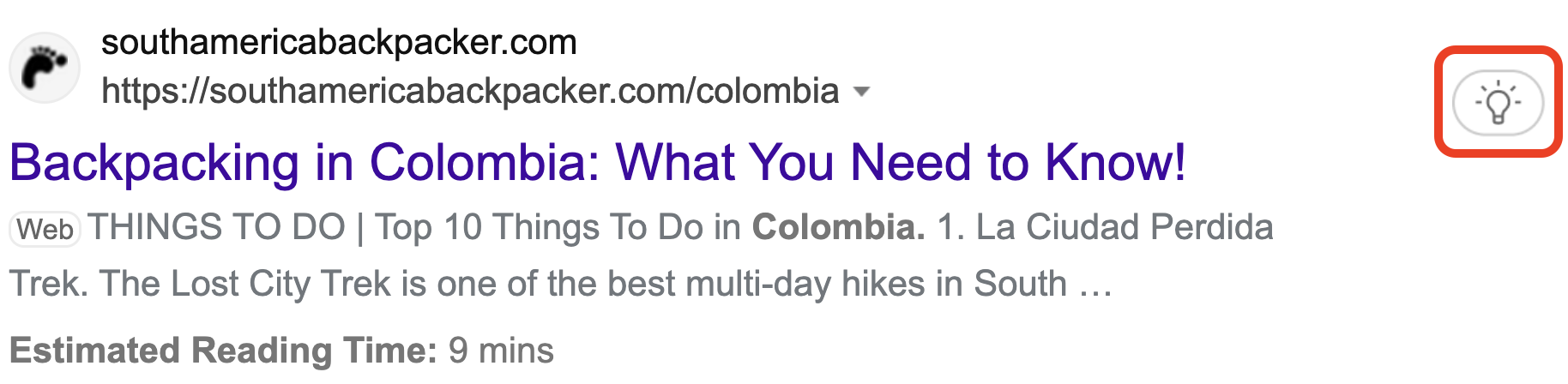}}
        \caption{}
        \label{bing_bulb}
        \Description[Bing search]{The search context provided by Bing can be accessed by clicking on the light bulb present along each search result as shown in the figure.}
    \end{subfigure}\\
    \hspace{-0.5cm}
    \begin{subfigure}[b]{0.34\textwidth}
        \centering   
        \setlength{\fboxsep}{0.1pt} 
    \setlength{\fboxrule}{1pt}
    \fbox{\includegraphics[width=1\textwidth]{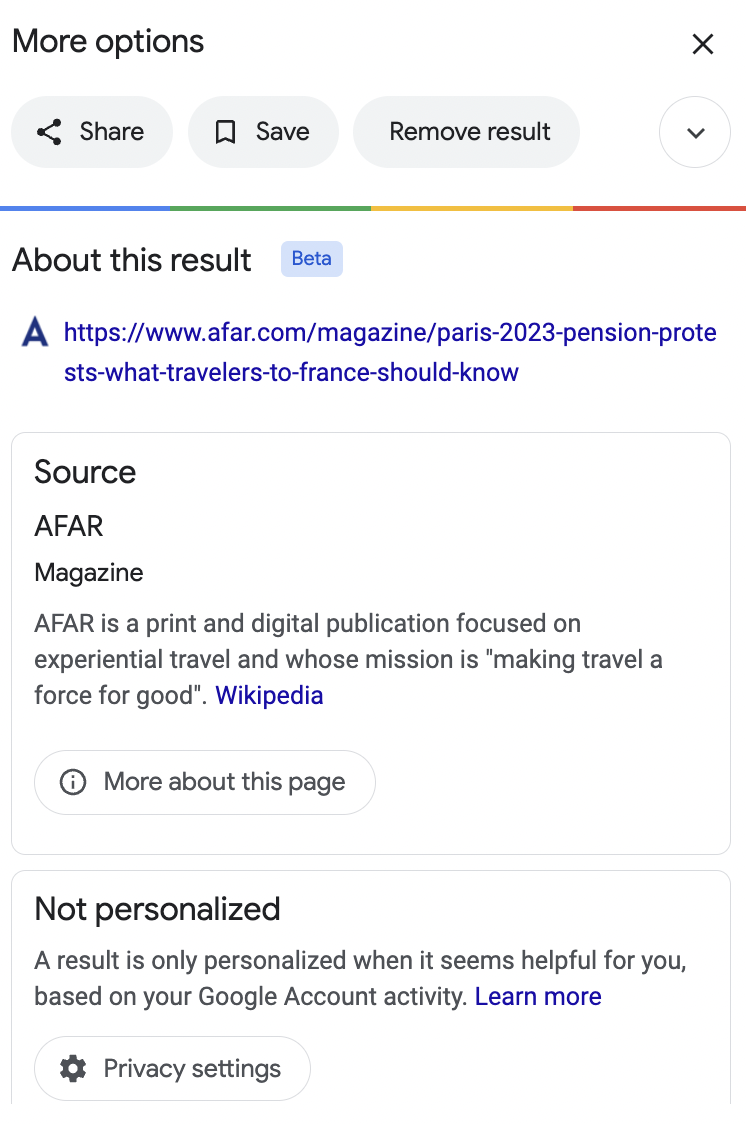}}
        \caption{}
        \label{g1}
        \Description[Search explanations provided by Google]{Figure shows the search result explanations and additional context presented by the Google search engine alongside each search result.}
    \end{subfigure}
    \hspace{0.3mm}  
    \begin{subfigure}[b]{0.342\textwidth}
        \centering      
         \setlength{\fboxsep}{0.1pt} 
    \setlength{\fboxrule}{1pt}
    \fbox{\includegraphics[width=1\textwidth]{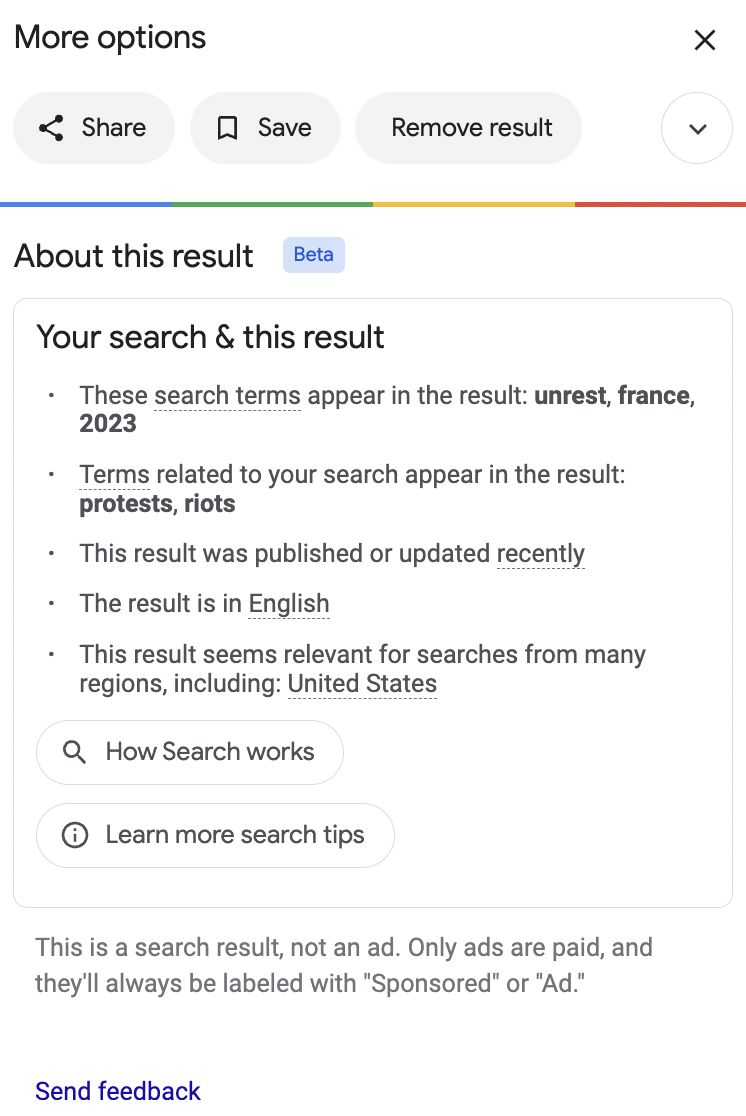}}
        \caption{}
        \label{g2}
        \Description[Search explanations provided by Google]{Figure shows the search result explanations and additional context presented by the Google search engine alongside each search result.}
    \end{subfigure}
     \hspace{0.3mm} 
\begin{subfigure}[b]{0.31\textwidth}
        \centering
         \setlength{\fboxsep}{0.1pt} 
    \setlength{\fboxrule}{1pt}
\fbox{\includegraphics[width=1\textwidth]{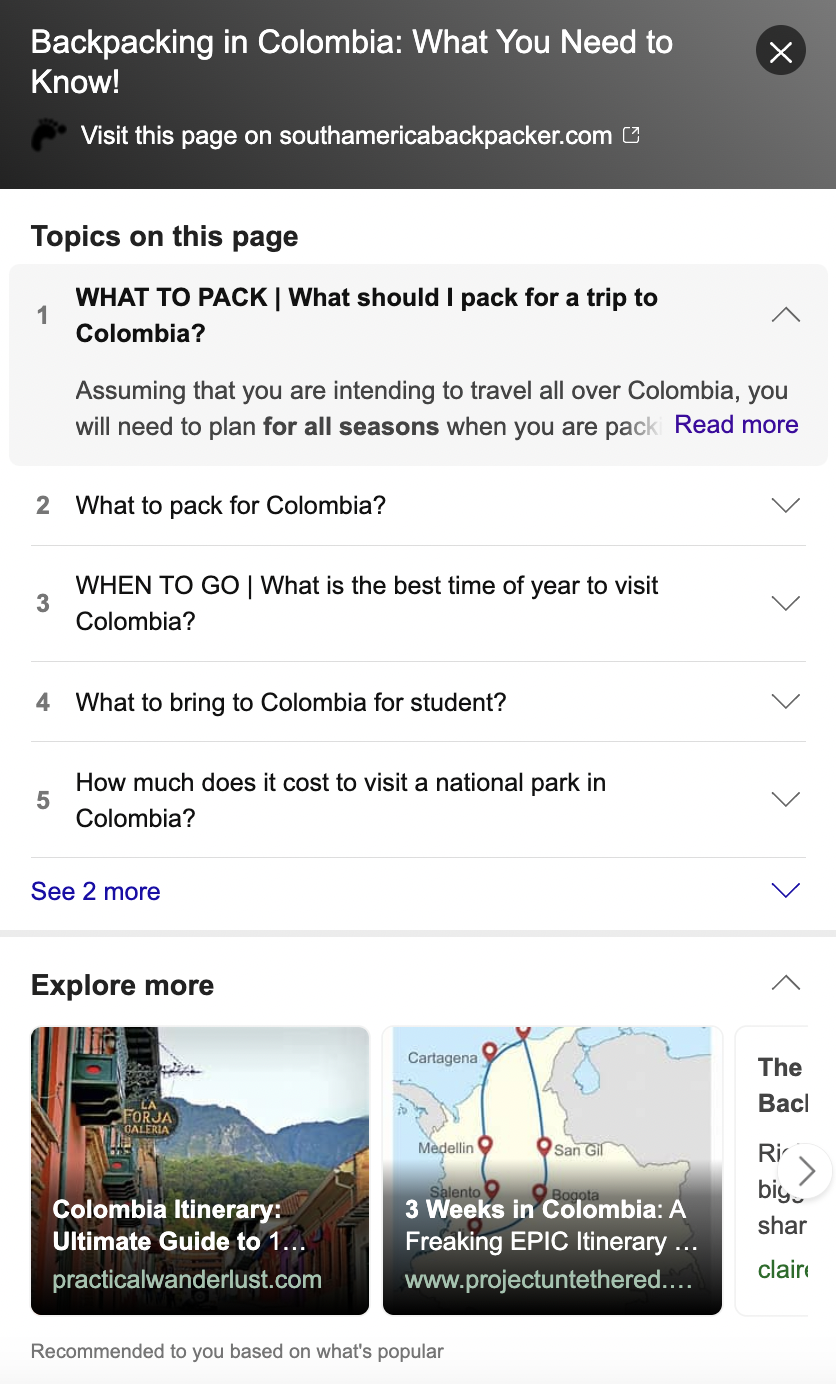}}
\caption{}
              \label{bing}
              \Description[Search context provided by Bing]{Figure shows the additional details that Bing provides along with each search result.}
    \end{subfigure}
    \caption{Figures \ref{g1} and \ref{g2} demonstrate the search result explanations and additional context presented by the Google search engine alongside each search result. These details can be accessed by clicking on the three vertical dots next to each search result as shown in Figure \ref{g_dots}. The provided data includes various features such as the option to remove a result and provide feedback on the search result page, a Wikipedia link for accessing supplementary information about the source, a link to privacy/personalization settings, an explanation for the appearance of the specific search result, and a link that directs users to a page offering a broad overview of how search results are curated by search engines. On the other hand, Figure \ref{bing} illustrates the search context provided by Bing. This can be accessed by clicking on the light bulb present along each search result as shown in Figure \ref{bing_bulb}. It encompasses details about the source (if available on Wikipedia/Encyclopedia), a preview of relevant topics on the webpage, and additional recommendations associated with the search result.}
\end{figure*}

\begin{table*}[]
\begin{tabular}{l|l|l}
\hline
\textbf{Feature} & \textbf{Google} & \textbf{Bing} \\ \hline
Source information & present & present \\ \hline
Web page preview & not present & present \\ \hline
Search explanations & present & not present \\ \hline
Option to provide feedback & present & not present \\ \hline
Link to privacy and personalization settings & present & not present \\ \hline
Popular recommendations related to search result & not present & present \\ \hline
\end{tabular}
\caption{Table illustrating the comparison of key features shown by Google and Bing alongside each search result. }
\label{diff}
\Description[Additional details provided by Google and Bing]{Table illustrating the comparison of key features shown by Google and Bing alongside each search result.}
\end{table*}

\subsection{Desired characteristics of search result explanations}

Table \ref{characteristics} provides a summary of the desired characteristics of search result explanations as expressed by the participants. Their perspectives highlighted the importance of concise explanations, actionable guidance, indicators of credibility and reputability, and the inclusion of viewpoints in the explanations. Participants recognized the ``\textit{overwhelming nature of internet searches [that]  require sifting through [large amounts of] information}'' (P11), and expressed a strong desire for concise explanations that provide quick access to relevant information. As  P7 expressed---``\textit{To be really helpful, information [in the explanation] would need to be a very quick read}''.  Second, participants expressed the need for explanations to be actionable, i.e. they assist users with either selecting the search result or guiding the search process. Some participants expressed interest in having user experience features that allow them to change the order of search results based on specific criteria. For example, P7 suggested the inclusion of a button that enables users to adjust rankings based on their preferences. Participants also emphasized the presence of indicators of credibility and reputability of search results in the explanations. Furthermore, participants stressed the need for explanations to indicate the viewpoints of the search results. Understanding the perspective or stance of the source can help users assess the alignment of their own views and make more informed judgments about the information presented. As P1 mentioned, ``\emph{I would want to know who is teaching me this thing... make sure that their views aligned with mine}''.

\section{RQ3: Users' perception of search explanations and search context  provided by search engines}

During the interviews, participants were shown the search result explanations and additional context provided by Google (Figures  \ref{g1} and \ref{g2}) and Bing (Figure \ref{bing}) search engines. {Table \mbox{\ref{diff}} lists the key features shown by Google and Bing alongside each search result. Bing provides additional context about webpage source, and webpage preview, and lists popular recommendations related to the search result. On the other hand, Google provides source information, search explanations, links to personalization and privacy settings, and an option to provide feedback about the search result.} Interestingly, none of the participants were aware of the existence of these additional data accompanying the search results. The participants expressed diverse perspectives on these features, offering insights into their perception and usefulness. We discuss the emerging themes below.

\subsubsection{Option to access additional information is valued:} All participants acknowledged the value of having the option to access additional information about the search result {in both Bing and Google search engines}. As P7 noted, ``\textit{That gives a person the searcher the option of taking the extra time to learn the extra information... and that's always a good thing to have}''. Participants also appreciated the concise presentation of the information presented by search engines, as one participant mentioned, ``\textit{I like that it's concise and doesn't show you everything you want to see..you have to click somewhere else, which I do like because it's like it makes it less cluttered I guess}'' (P6). 

\subsubsection{Broad and obvious explanations are not deemed useful:} Many participants did not find the content of the search explanations, {currently only provided by Google}, particularly useful, deeming it to be either obvious or too broad. P5 provided an example stating that their getting a specific result due to their location in the USA is not very helpful---``\textit{I don't think the result in the region would be particularly helpful, especially if it's as broad as the United States}''. P2 concurred, expressing, ``\textit{I feel like a lot of this information doesn't necessarily help me out with some of those qualities that we were talking about..like the trust, the credibility, the efficiency}''.

\subsubsection{Contestability is desired in search:} {The majority of participants appreciated the functionality provided solely by Google to remove search results and offer feedback.}. They saw this feature as a means to contest and influence the search results, granting them a voice in determining the relevance and accuracy of the displayed information. Participants appreciated the agency and empowerment that came with being able to shape the search results according to their preferences and needs. One participant emphasized the significance of user input, expressing the necessity of having such a feature---``\textit{I feel like it's something that's necessary to have something like this because if they did not include like a send feedback option, it would feel sort of wrong in a way, it would feel more so that like Google is saying, we know what's right, we know what's best, and you know, you don't have a say in that}'' (P6). 

\subsubsection{Privacy and personalization settings:} {Participants found the privacy and personalization settings provided only by Google to be helpful.} The majority of them were not aware that they could change their privacy settings to alter the way Google collects their personal information to personalize the search results for them. Participants suggested that instead of having the privacy settings within the search explanations pop up, ``\textit{it should be easily accessible and visualizable on the search result page itself}'' (P3).

\subsubsection{Website previews increase search efficiency: } All 12 participants highlighted the usefulness of the website preview feature offered only by Bing that provides an overview of the website's content. Participants expressed a preference for this feature as it allowed them to get a ``\textit{whole vibe of the website without visiting it directly}'' (P2) and ``\textit{jump to the parts that are relevant to one's specific search}'' (P5). 

\subsubsection{Significance of recommendations along with search results:} Bing offers additional search result recommendations alongside the primary search results. Participants found them helpful especially when exploring a topic or seeking specific pieces of information. One participant stated, ``\textit{I think that's really helpful when you're doing something that you're maybe trying to learn more about a topic and you don't know where to start... you can kind of follow it down a rabbit hole to something more specific}'' (P5). 

\subsubsection{Source information enhances trust:} The majority of participants expressed their appreciation for the additional information about the source {presented by both Bing and Google}, underscoring its role in bolstering the perceived reliability of search results---`` \textit{I think linking me to Wikipedia can help me get like a broader understanding of the source with like a decent amount of trust in the results I'm seeing there.}'' (P9).

\subsubsection{Need to improve discoverability of explanation features:} For both search engines,  participants found it challenging to locate these features, expressing a desire for improved discoverability and accessibility. For example, one participant noted, ``\textit{I like that it's there, I wish it was a little bit more transparent so you didn't have to click on a bubble and like three dots get to that thing}'' (P5). On similar lines, P11 added, ``\textit{I mean it's so deeply layered... I don't know if I'd ever be able to find it}''.





\section{Discussion}
In this study, we interviewed non-technical individuals to investigate the situations when users question search results, the search objectives where users need explanations, and the desired characteristics and benefits of search explanations. Our findings have implications for the design of search engines and search explanations. We discuss them below.

\subsection{Users' needs in evaluating search results}
{Before introducing the concept of explanations, we prompted users to articulate the additional information they believe search engines should offer to aid in their evaluation of results. Interestingly, none of the users explicitly mentioned explanations. Instead, their responses highlighted a desire for efficiency-enhancing features and credibility indicators that the users believed would improve their search experience. Throughout interviews, participants consistently expressed their need for swift decision-making regarding whether to click on a search result. To meet this demand, search engines have the opportunity to offer several efficiency-enhancing features. For example, introducing website preview functionality that activates when users hover over a search result. Search engines can also adopt Bing's preview feature that lists the headings present on a webpage as topics with hyperlinks, giving users a quick and informative overview of the page's content along with the ability to directly navigate to the desired section of the webpage.  While efficiency is desired by the majority of users, it can sometimes inadvertently lead users to not critically evaluate search results. 
Recognizing this potential pitfall, initiatives like Microsoft's Search Coach \mbox{\cite{Introduc70:online}} have been designed to encourage reflective browsing by encouraging users to pause, think, and seek external information about the source or facts presented on the webpage, and also offer tips that train users on how to craft effective search queries \mbox{\cite{Introduc70:online,TheChatG73:online}}.} 

{In addition to efficiency, participants also expressed a need for information regarding the credibility of search results. They proposed credibility indicators such as authoritativeness, website views, and user-generated reviews. Search engines can incorporate credibility into their design in various ways. One approach could involve enabling users to leave website reviews, fostering a collective, crowdsourced credibility assessment system, akin to initiatives like Birdwatch on the X platform (formerly Twitter), where users collaboratively evaluate information credibility \mbox{\cite{wojcik2022birdwatch,allen2022birds}}. Additionally, Google's ``Your Money or Your Life'' (YMYL) concept currently employs manual evaluation by Search Quality Raters (SQR) to ensure higher standards of expertise, authoritativeness, and trustworthiness for crucial topics related to finance, health, law, and other sensitive areas \mbox{\cite{WhatareG92:online}}. Search engines can display SQR evaluations as explanations alongside search results, aiding users in assessing content quality. However, there are also several potential limitations to indicating credibility. First, certain topics may inherently lack a definitive truth value (e.g. philosophical debates, dietary recommendations, etc.). Second, crowdsourced methods of credibility assessments could also be impacted by user bias \mbox{\cite{allen2022birds}}. Despite these limitations, there are certain facts whose truth values are known and established by scientific sources (e.g. vaccines do not cause autism) for which search engines can provide valuable context. Particularly when search engines have guidelines related to the credibility of content (e.g. Google's SQR), offering insights into the underlying policies and providing additional context about the decisions would be helpful for the users.}

\begin{figure*}[]
  \centering
      \includegraphics[scale=0.3]{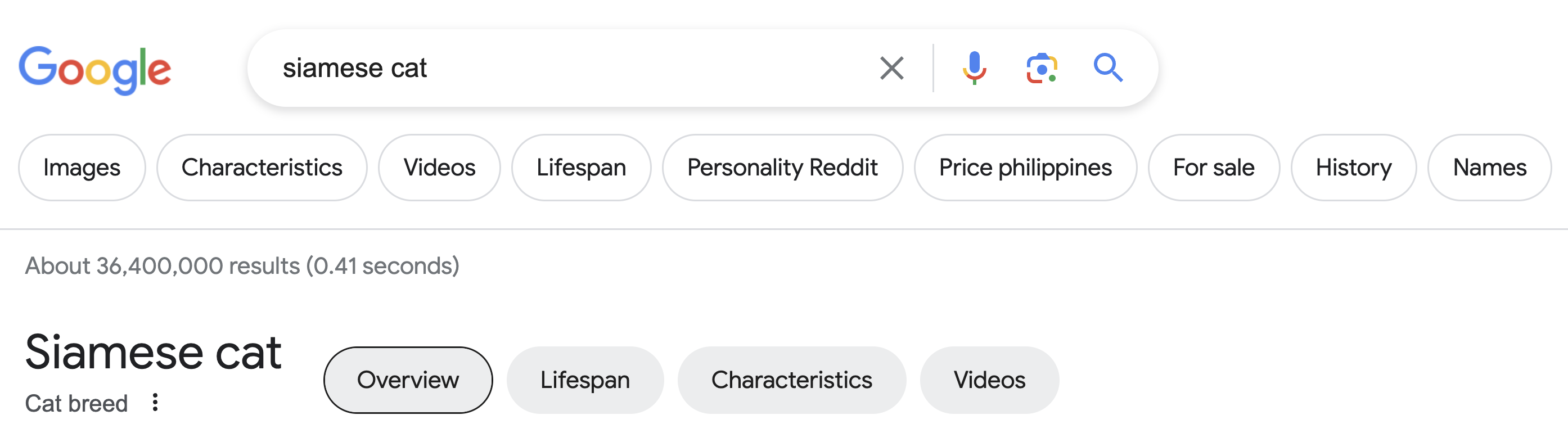}
  \caption{Google's topic filters feature introduced in 2023 \cite{Googlero39:online}
  }
  \label{google_topics}
  \Description[Google's topic filters]{Figure illustrating Google's topic filters feature introduced in 2023, showcasing a range of topics associated with the searched topic.}
\end{figure*}


\subsection{If and when users need search explanations}
\subsubsection{When explanations are not needed} {During the study, when users were not acquainted with the concept of explanations, they revealed that they don't think too deeply about the search results in their daily lives. Additionally, they only questioned content that appeared irrelevant, inappropriate, or sponsored.    This indicates that users generally are not concerned by how search engines curate content as long as their needs are met. Notably, our study also highlighted users trust in the results shown by search engines,  a phenomenon well-documented in prior literature \mbox{\cite{pan2007google}}.}


{Once acquainted with explanations, participants further expressed that they would not need explanations for searches where the goal is to find a single, unambiguous answer. Search engines typically respond to such requests in a featured snippet or answer box. Despite users' belief in the accuracy of search snippets or answer boxes, they do not always contain accurate information. For example, while there is no evidence that any American president has been a member of the racist cult Ku Klux Klan,  Google of 2017 incorrectly returned names of American presidents in response to the query ``presidents in the klan'' \mbox{\cite{Google’s76:online}}.  This underscores the need for search engines to exercise caution when highlighting responses to such queries, prioritizing the provision of credible and accurate information in search snippets and answer boxes. Furthermore, this finding has broader implications for the future of search. With the rise of generative AI, search engines have begun showcasing chat-bot-generated responses as the topmost result, emphasizing the potential time-saving benefits of using chatbots for search queries. As the use of chatbots becomes prevalent, there is a possibility that users might even get less inclined to check up on the bots' answers independently. This poses an intriguing avenue for future research---to explore how users respond to chatbot-generated answers and assess their impact on the search experience. It's also a call for search engines to carefully consider the implications of this evolving landscape and the potential consequences for user trust and information accuracy.}

\subsubsection{When explanations are needed} {As users familiarized themselves with the concept of explanations, they acknowledged their significance but in limited contexts. Users revealed that they 
would desire explanations when the task requires deep engagement and when they have the time to spend on the search. Users also said that they require explanations for topics crucial to well-being, such as searches involving medical inquiries. Explanations are also deemed useful for searches where the objective is to find real-world services or products. The significance of explanations in these situations was underscored by the anticipation of interacting with products or visiting specific places. For all the aforementioned search objectives, users emphasized their preference for concise and easy-to-read explanations. This preference indicates that users want to spend minimal time engaging with explanations, highlighting the importance of providing information in a quick and accessible format. It's noteworthy that despite the significant research conducted in the past decade on creating explanations \mbox{\cite{guidotti2018survey}}, some of which are complex and contain detailed insights into the functioning of automated systems, our findings suggest that while such explanations have value for understanding how automated systems work, non-technical users may not fully engage with them.}

\subsection{Design implications}
{For search objectives where users find explanations helpful, they disclosed various potential benefits and expressed preferences for specific characteristics in explanations. Each of these aspects has significant implications for the design of explanations and search engines, which we explore in the following sections.}

\subsubsection{Content Categorization and Exploration} Participants expressed a need for explanations to indicate categories of content, particularly in searches where the objective is either to obtain answers to open-ended questions or to comprehensively explore a topic. They revealed that such categories would enable them to swiftly identify redundant information and concentrate on acquiring new knowledge. Search engines have taken initial steps in addressing this need, exemplified by the introduction of topic filters by Google since our interviews were conducted. For instance, when searching for ``Siamese cat'', users can access specific categories such as breed characteristics, lifespan, history, and more, as shown in Figure \mbox{\ref{google_topics}}. Search engines can further enhance topic filters by allowing user customizations including the ability to specify the level of categorization granularity, offering interactive previews that provide concise summaries of category results, and incorporating mechanisms for user feedback.


\subsubsection{Guiding the Search Process} {Participants also desire explanations to guide the search process by aiding in refining the search queries. While contemporary search engines currently offer snippets with each result, containing highlighted words to indicate the alignment of the search query with the result, as well as alternate query suggestions, there are opportunities for further enhancements. Search engines could provide more transparency about query transformations, explaining how the search query was ultimately processed. Additionally, they can educate users on how to use advanced search features like quotation marks or boolean operators to articulate their information needs more effectively. Moreover, they can go beyond standard query suggestions by presenting alternative query results and showcasing side-by-side comparisons to help users find the most relevant query.}

\subsubsection{Indication of Viewpoints in Explanations}{ Participants sought explanations that indicate the viewpoints being presented and suppressed in search results. Suggested approaches included displaying diversity metrics and incorporating banners to explain suppressed viewpoints. These features aim to provide users with a more comprehensive understanding of the perspectives covered in search results. Previous research has explored several intervention designs to indicate the diversity of search results \mbox{\cite{munson2013encouraging,mattis2022nudging,paramita2023towards}}, which have proven beneficial in making users more aware of potential biases in search. Incorporating such features in explanations could further enhance users' awareness of the diversity of perspectives represented in search outcomes.}

\subsubsection{Actionable explanations} {
Interviews underscored users' desire for explanations that not only elucidate search engine processes or provide supplementary information but also offer actionable insights. Notably, participants appreciated Google's search explanations that informed them about personalized search results, accompanied by a convenient link for adjusting privacy and personalization settings. This finding emphasizes the pivotal role of explanations in empowering users to take tangible actions, enhancing their control over the search experience, and ensuring alignment with their preferences and requirements. }

\subsubsection{Contesting Search Results and User Agency} {Another crucial aspect that users expect from explanation interfaces is the ability to contest search results. The majority of participants expressed appreciation for Google's functionality that enables them to remove a search result and offer feedback. This feature empowers users to take an active role in shaping their search experience, ensuring that the results align with their expectations and requirements. However, many users were unaware of its existence. Given the usefulness of this feature, it becomes essential for search engines to increase users' awareness of its presence and functionality. This can be achieved through intuitive interfaces, user-friendly tutorials, or proactive prompts that encourage users to explore and utilize these features. }

\section{Limitations and future opportunities}\label{limitation}
While our work offers valuable insights, it is not without limitations.  {Our study had a limited sample size and focused on English-speaking ``non-technical'' individuals residing in the United States, with a majority being White/Caucasian. Consequently, our findings do not comprehensively capture user expectations and preferences across diverse regions, and ethnic backgrounds. Regional variations can significantly shape users' expectations from search results and in turn search explanations by influencing the specific nuances, cultural context, and contextual relevance they seek from search results. Similarly, the racial background of users can also have an impact. Users from specific racial backgrounds may feel marginalized or dissatisfied if they consistently observe under-representation or stereotyping in search results. In our study, our primary focus was on gaining a broad understanding of the helpfulness of explanations in specific search scenarios, laying the groundwork for more detailed investigations into these influential factors in the future.}

{We did not account for personalization in our study. For instance, consider location-based personalization, where, when users search for `Italian restaurants', the relevance and practicality of information becomes crucial. Users are more likely to find search results satisfactory when aligned with their local context, which might diminish the need for detailed explanations. While personalization can significantly influence users' satisfaction with search results and, subsequently, their need for explanations, we acknowledge this as an area for future studies to explore in depth.}

{Another limitation of our study stems from our definition of ``non-technical'' users, which relied on self-reported technical competence and educational degree. However, we recognize that the term ``non-technical'' encompasses a wide spectrum of characteristics, extending beyond education degrees to include factors such as digital literacy levels, critical-thinking skills, and awareness of information disorders such as bias and misinformation. Unfortunately, we did not collect data on these nuanced aspects. As a result, our study might not fully represent the complexities inherent in the ``non-technical'' user group, limiting the ability to draw robust conclusions about the specific needs and expectations of users within this category.  For instance, users with high critical-thinking skills and advanced digital literacy might have sought detailed explanations for all search objectives, desiring more granular insights into search engine processes within explanations. Additionally, our study focused exclusively on Google and Bing search engines, omitting others such as Yandex, YouTube, DuckDuckGo, etc. Future studies should aim to address these gaps for a more comprehensive understanding and generalizability across diverse populations, regions, and search engines.}

Our study interviews were conducted in September 2022, before the widespread incorporation of Language Models (LLMs) in search engines. These LLMs have disrupted the search dynamics by providing single answers directly at the top of search results. However, our findings still hold value as they highlight the importance of providing explanations and citations to enhance user trust in search results. There is a future opportunity to explore the role of search explanations within the context of LLMs. Understanding how explanations can be effectively integrated with these models is a crucial area for further research.
Additionally, our study's goal is not to delve into the kind of explanations that would be most helpful in different scenarios and contexts. Future research can investigate the effectiveness of various explanation formats, such as visual or interactive formats, and explore different types of explanations, like counterfactual or contrastive explanations, to enhance user understanding and engagement with search results.

\section{conclusion}
In this study, we conduct a qualitative study to understand users' needs from search explanations.  We investigate if explanations are helpful, identify the specific search objectives in which they are beneficial, and elucidate the perceived advantages of explanations across various contexts. We find that users value search explanations for complex or critical tasks. They do not find explanations helpful when they are seeking short unambiguous answers.  They expressed a preference for concise, actionable explanations that could assist them in refining their search queries and gaining a deeper understanding of search result perspectives. Additionally, the study shed light on the perception of existing search engine features, with participants appreciating the ability to contest search results and preview search result web pages. Overall, this research contributes to the ongoing discourse on search engine transparency and offers valuable recommendations for enhancing user engagement and trust in search results, thereby improving the overall search experience.

\bibliographystyle{ACM-Reference-Format}
\bibliography{sample-base}

\appendix

\end{document}